\documentstyle{mn}
\def\rosat{{\it ROSAT}}

\def\cpd{ CPD\,-59$^{\circ}$\,2635}

\begin{document}

\title[ CPD -59$^{\circ}$\,2635: A new Double-Lined O type Binary ]
      {Optical spectroscopy of X-MEGA targets\\ 
I. CPD -59$^{\circ}$\,2635: A New Double-Lined O type Binary in 
the Carina Nebula}

\author[J. F. Albacete Colombo, et al.]
 {J.F. Albacete Colombo$^1$\thanks{Visiting Astronomer,
 CASLEO, operated under agreement between CONICET and National Universities
 of La Plata, C\'ordoba and San Juan, Argentina.}
 \thanks{Fellow of CIC, Prov. de Buenos Aires, Argentina.},
  N.I. Morrell$^1${\Huge $^{\star}$}
 \thanks{Visiting Astronomer, CTIO, NOAO, operated by AURA Inc., for NSF.}
 \thanks{Member of Carrera del Investigador Cient\'{\i}fico,
 CONICET, Argentina.},
  V.S. Niemela$^1${\Huge $^{\star}$} 
 \thanks{Member of Carrera del Investigador Cient\'{\i}fico, CIC, Prov. 
 de Buenos Aires, Argentina.}
 and M.F. Corcoran$^{2,3}$ 
\\
$^1$ Facultad de Ciencias Astron\'omicas y
Geof\'{\i}sicas, Universidad Nacional de La Plata,
Paseo del Bosque S/N, 1900 La Plata, Argentina\\
$^2$ Universities Space Research Association, 7501 Forbes
Blvd, Ste 206, Seabrook, MD 20706, USA\\
$^3$ Laboratory for High Energy 
Astrophysics, Goddard Space Flight Center, Greenbelt MD 20771, USA\\
}
\maketitle
\begin{abstract}
Optical spectroscopy of CPD -59$^{\circ}$\,2635, one of the O-type stars 
in the open cluster Trumpler 16 
in the Carina Nebula, reveals this star to be a double-lined 
binary system.
We have obtained the first radial velocity orbit for this system,
consisting of a circular solution with
 a period of 2.2999 days and semi amplitudes of 
208 and 273~km~s$^{-1}$. This results in minimum masses of 15 
and 11~M$_{\odot}$ for the binary components of \cpd,
which we classified
as O8V and O9.5V, though spectral type variations of the order of 1 
subclass, that we identify as the {\it Struve-Sahade effect},
seem to be present in both components. 
From ROSAT HRI observations of \cpd\ we determine a luminosity ratio 
log$(L_{\rm x}/L_{\rm bol}) \approx -7$, which is similar to that
observed for other O-type stars in the Carina Nebula region.
 No evidence of light variations is
present in the available optical or X-rays data sets.

\end{abstract}


\section{Introduction}

\begin{figure*} 
\vspace{85mm}
\caption{Optical field around CPD $-59^{\circ}$\,2635 (left) from
digitized sky survey and \rosat\ HRI image (right).  The fields of
view are $6'\times 6'$.}
\includegraphics{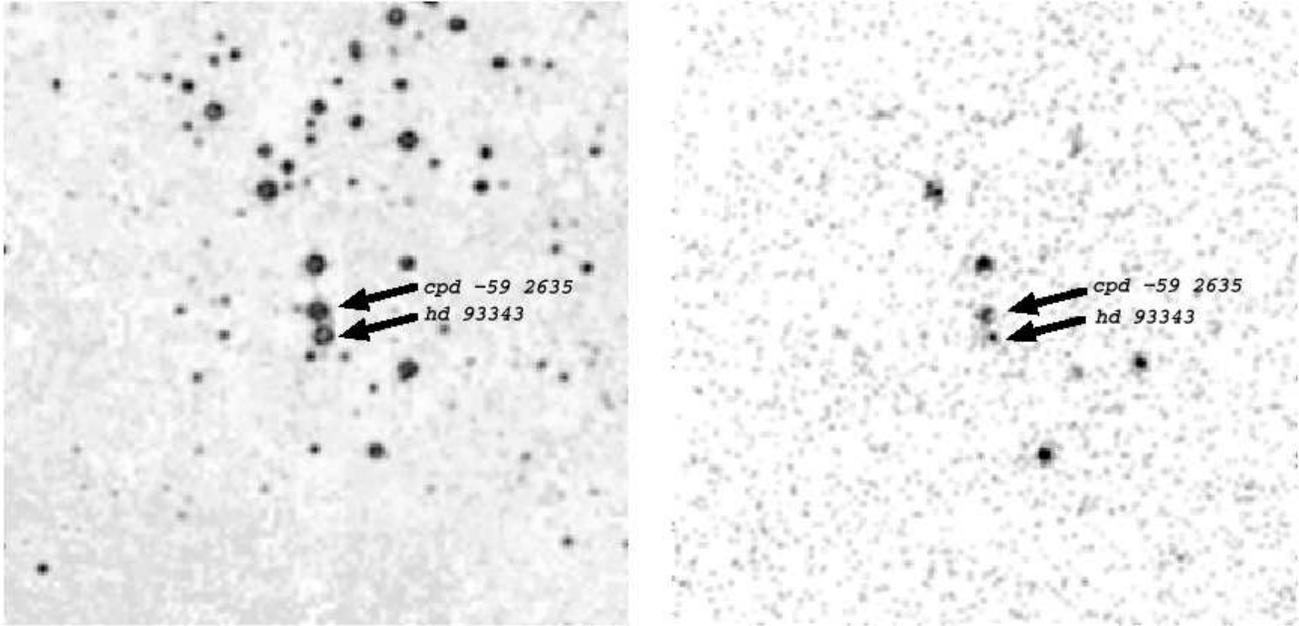}
\label{hri_dss_cmp.eps}
\end{figure*}

The very young Carina Nebula region contains several open clusters with
a rich population of O-type stars. Among them, Trumpler 16 is one of 
the most conspicuous. One of its members, 
CPD\,-59$^{\circ}$\,2635 (V=9.27, $\alpha_{2000}$=10$^{\circ}$ 45$'$ 12.78$''$, 
$\delta_{2000}$=-59$^{\circ}$ 
44$'$ 46.6$''$; Massey \& Johnson, 1993) 
has been observed in the context of the 
international X-Mega campaign, which involves optical 
spectroscopy of OB stars showing X-ray emission on ROSAT-HRI 
images {(Corcoran 1999)~\footnote{(See http://lheawww.gsfc.nasa.gov/users/corcoran/xmega/ for a more comprehensive description of the 
XMega campaign).}}
CPD\,-59$^{\circ}$\,2635 is one of the OB stars in the 
neighborhood of $\eta$ Carin\ae\ also detected as a bright  X-ray source.
Figure~\ref{hri_dss_cmp.eps} shows an X-ray image centered on 
CPD\,-59$^{\circ}$2635 
obtained through combination of 3 deep \rosat\ HRI pointings (see below)
along with the optical field from the Digitized Sky Survey.

Because a massive binary system could influence its emergent X-ray flux
as a result of colliding stellar winds (e.g. Chlebowski \& Garmany, 1991),
it is important to verify the frequency of close multiple 
systems among the Carina OB stars which are detected as
 X-ray sources. We wonder,
for example, if wind collision might be
the physical reason that makes CPD\,-59$^{\circ}$\,2635
brighter in X-rays than its close neighbor
HD~93343 
 very similar in both visual brightness and spectral type 
(see Fig.~\ref{hri_dss_cmp.eps}). 

CPD -59$^{\circ}$\,2635 has received different designations in the literature. 
The IDS  catalogue (Jeffers et al. 1963) refers to it 
as IDS 10452 S 5946  as possibly forming a visual binary with HD~93343, 
of similar V magnitude, about 14'' to the South. 
Stephenson \& Sanduleak (1971)  gave to CPD~-59$^{\circ}$\,2635 
the number 1872 in their catalogue of ``Luminous Stars in the Southern
Milky Way'' (LSS), but misleadingly providing the
cross-identification as ``HD 93343?''.  Feinstein et al. (1973)
assigned to CPD\,-59$^{\circ}$\,2635 the number 34, 
among the probable members of the open cluster Trumpler~16. 

\section{Observations}

Our observations consist of optical spectrograms of CPD -59$^{\circ}$\,2635,
 obtained during 1984 at Cerro Tololo Interamerican Observatory (CTIO), 
Chile, and between 1997 and 2000 
at Complejo Astron\'omico El Leoncito (CASLEO).

The first set of 11 observations was obtained in March 1984 
at CTIO using the Carnegie Image Tube Spectrograph (CTIS) attached 
to the 1-m Yale telescope. 
These spectrograms, covering a wavelength range from 3900 to 4900 \AA\
at a reciprocal dispersion of 43 \AA\ mm$^{-1}$ 
were widened  to 1~mm and secured on Kodak IIIa-J baked emulsion.
A He-A lamp was used as a comparison source.
These photographic spectrograms were 
digitized with a Grant micro-densitometer at La Plata Observatory, 
and subsequently analyzed with IRAF \footnote
{Image Reduction and Analysis Facility,  distributed by NOAO,
operated by AURA, Inc., under agreement with NSF.} routines.

Spectral CCD images of CPD -59$^{\circ}$\,2635 were obtained 
at CASLEO observatory, between January 1997 and June 2000
with the 2.15-m Jorge Sahade telescope, mainly as part of the observations
for the XMega campaign.
We used a REOSC echelle Cassegrain spectrograph 
and a Tek 1024 $\times$ 1024 pixels CCD as detector to obtain
27 spectra in the wavelength range from 
3500 to 6000~\AA~ at a reciprocal dispersion of 0.17 \AA\ px$^{-1}$ 
at 4500~\AA.
The S/N of these data is $\sim$ 110 (although it changes, of course,
with position within each echelle order).

Four additional observations were obtained at CASLEO with a 
Boller \& Chivens (B\&C) spectrograph attached to the 2.15m telescope, using a  
PM 516 $\times$ 516 pixels CCD as detector, 
and a 600 l mm$^{-1}$ diffraction grating, yielding a
reciprocal dispersion of 2.5 \AA\ px$^{-1}$.
These spectra cover the spectral range from $\sim$ 3800 to 4900 \AA\ ,
and their S/N is $\sim$300.
One more spectrum of \cpd\ was obtained at CASLEO with the REOSC spectrograph
in its simple dispersion mode, using a 600 l mm$^{-1}$ grating and the
Tek 1024 $\times$ 1024 CCD as detector, at a resulting reciprocal
 dispersion  of 1.8 \AA\ px$^{-1}$. The central wavelength of this
observation is 4700 \AA\ and the corresponding S/N is $\sim$ 300.
 
The usual series of bias, flat field and dark exposures were also
secured during each observing night for every CCD data set.
The CCD images were processed and analyzed with IRAF routines at 
La Plata Observatory.

\section{Results and their Discussion}

\subsection{Radial Velocity orbit of CPD\,-59$^{\circ}$\,2635}

\begin{figure}
\vspace{75mm}
\caption{He{\sc ii} 4686~\AA\ absorption in the spectrum of 
 CPD~-59$^{\circ}$\,2635 at different observing dates, 
showing the doubling
 of spectral lines.}
\includegraphics{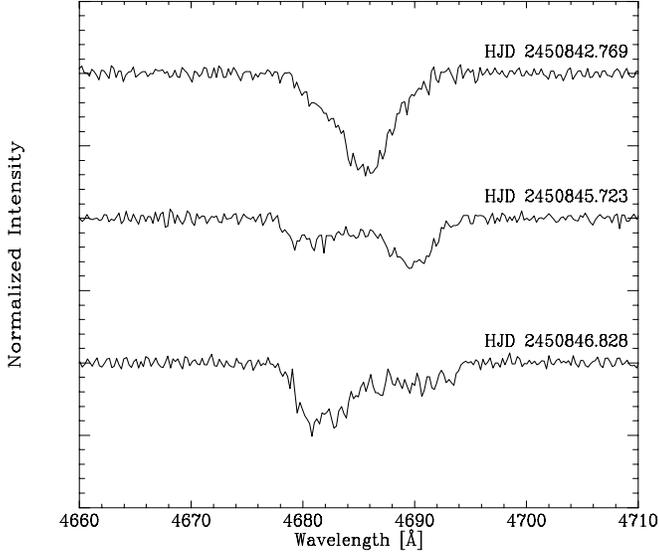}
\label{dl4686}
\end{figure}

A first inspection of our high resolution echelle spectrograms revealed 
 double lines present in some of them, indicating that 
CPD -59$^{\circ}$\,2635
was probably a double-lined spectroscopic binary.
Figure~\ref{dl4686} shows the behavior of He {\sc ii} 4686 \AA\ line in 
echelle spectra of CPD\,-59$^{\circ}$\,2635, obtained 
at different observing dates.

Radial velocities were determined from our 
spectra of \cpd,
fitting Gaussian profiles to the absorption lines.

We used for radial velocity determination several lines of He{\sc i} along
with He{\sc ii} 4686 \AA\, which appear as the least affected by pair blending.
The Pickering (4-n) series of He{\sc ii} absorptions of the two binary
components are not well resolved in our spectra, these lines being broader than
other He{\sc ii} and He{\sc i} absorptions. 
 We interpret this fact as indicative
of higher pressure broadening acting in the region where these lines form,
which must be deeper in the atmosphere than the formation region of
He{\sc i} lines. As a consequence, these (4-n) series lines are more
seriously blended than the other absorptions in the spectrum of \cpd\
and we decided not to include them in the average  radial velocities
presented in Table~1. 
We also derived radial velocities from our lower resolution, but higher
S/N CCD spectra and from our digitized photographic plates.
The radial velocities were computed as unweighted mean values of individual
velocities determined for each spectral line.

The journal of our radial velocity observations is presented in Table 1.
In successive columns, we quote the Heliocentric Julian Date (HJD),
the corresponding orbital phase (as explained below),
 the measured average radial velocities for the primary and secondary
components, and their standard deviations (s.d.).
We identified the primary component as the one 
having deeper absorptions of He{\sc ii} lines.

\begin{table}
\caption{Radial-velocity measurements for CPD -59$^{\circ}$\,2635.}
\label{symbols}
\begin{tabular}{lcrcrc} 
\\
  HJD   & phase&Primary &s.d. & Secondary&s.d. \\ 
2400000+&$\phi$&$km\,s^{-1}$&$km\,s^{-1}$&$km\,s^{-1}$&$km\,s^{-1}$\\
\\
45776.826$\dag$ & 0.86 & 193      & 34  & -132      & 56 \\
45777.783$\dag$ & 0.27 & -214     & 56  & +256      & 41  \\ 
45778.731$\dag$ & 0.63 & +179     & 22  & -210      & 28  \\
45779.715$\dag$ & 0.11 & -138     & 25  & +160      & 10   \\ 
45780.748$\dag$ & 0.56 & +27      & 25  & ---       & --- \\
45781.644$\dag$ & 0.95 & +37      & 28  & ---       & --- \\
45782.730$\dag$ & 0.42 & -116     & 46  & +98       & 28 \\
45783.677$\dag$ & 0.84 & +175     & 25  & -223      & 61 \\
45784.682$\dag$ & 0.27 & -204     & 47  & +219      & 68 \\
45785.634$\dag$ & 0.62 & +181     & 27  & -210      & 30 \\ 
45786.623$\dag$ & 0.12 & -135     & 28  & +180      & 34 \\ 
50495.835 & 0.64 & +163     & 23  & -225      & 18  \\
50498.826 & 0.92 & +108     & 10  & -131      & 18 \\
50506.834 & 0.43 & -68      & 19  & +132      & 13  \\
50841.793 & 0.07 & -86      & 27  & +152      & 28  \\
50842.769 & 0.49 &  +3      & 9   & ---       & ---  \\
50843.727 & 0.91 & +132     & 18  & -143      & 16  \\ 
50844.718 & 0.34 & -168     & 10  & +236      & 29  \\
50845.723 & 0.77 & +205     & 11  & -271      & 37  \\
50846.828 & 0.26 & -202     & 27  & +270      & 21  \\
50847.649 & 0.61 & +150      & 10  & -184      & 35  \\
50848.779 & 0.10 & -133      & 12  & +189      & 24  \\
50850.724 & 0.95 & +86       & 17  & -101      & 18  \\
50851.655 & 0.35 & -162      & 17  & +231      & 24  \\
50852.828 & 0.89 & +177      & 19  & -191      & 18 \\
50859.789$\ddag$ & 0.61& +112 & 21  & ---      & ---\\
50861.781$\ddag$ & 0.76& +189 & 24  & -248     & 25  \\
50862.762$\ddag$ & 0.18& -192 & 18  & +243     & 33 \\
50868.714$\ddag$ & 0.77& +188 & 22  &  ---     & --- \\
51208.709 & 0.61 & +134      & 17  & -214      & 21 \\
51209.714 & 0.04 & -49       & 27  & ---       & --- \\
51210.727 & 0.47 & -36       & 18  &  ---      & ---   \\
51211.715 & 0.90 & +118      &  6  & -179      & 24 \\
51215.682 & 0.63 & +137      & 17  & -197      & 17 \\
51216.653 & 0.05 & -82       & 20  & +80       & 12 \\
51217.673 & 0.49 & -12       & 21  &  ---      & ---   \\
51218.632 & 0.91 & +109      & 32  & -149      & 37 \\ 
51712.448 & 0.62 & +136       & 20  & -193      & 30 \\
51712.478 & 0.63 & +179       & 56  & ---       & --- \\
51715.463 & 0.93 & +84        & 16  & -120      & 10 \\
51716.483 & 0.37 & -165      & 33  & 184       & 19 \\
51716.521 & 0.39 & -131      & 16  & 160       & 23 \\  
51718.555$\ddag$ & 0.27 & -220      & 59  & 220    & 31 \\
\end{tabular}

 \medskip
 NOTE: Orbital phases have been calculated with ephemeris of table 2.
 $\dag$ indicates the lower resolution observations obtained 
 with the CITS  , and $\ddag$ those obtained either with the B\&C spectrograph
or the REOSC spectrograph in simple dispersion mode. 
\end{table}

From the radial velocities listed in Table 1, it was already apparent 
that the orbital period of \cpd\ was of the order of a few days. 
A period search routine based on the modified Lafler \& Kinman (1965)
 method (Cincotta, M\'endez \& Nu\~nez 1995)
applied to all the radial velocity observations of the primary component
 of \cpd\ as listed in Table 1, 
yielded as the most probable period $P = 2.29995 \pm 0.00002$ days.
Initial orbital elements were estimated, leaving also the period as a free
parameter, resulting in an orbital solution of negligible eccentricity 
($e =  0.005 \pm 0.008 $) with no significative change in
 the orbital period.  We therefore considered the orbit to be circular,
and the above mentioned value of the period to be the most probable,
and proceeded to find the best fit for the remaining orbital parameters.
In order to avoid as much as possible pair blending effects, 
we computed the orbital elements of \cpd\ using only radial velocities
 derived from our high resolution observations of both binary components,
 obtained at the  orbital phase intervals  0.1 to 0.4 and 0.6 to 0.9, 
of the binary period.
 The resulting orbital elements are listed in Table 2.
  
\begin{figure*} 
\vspace{100mm}
\caption{Radial velocity orbit for CPD -59$^{\circ}$\,2635.
The meaning of the symbols is as follows: 
{\it filled and open} symbols refer to the primary and secondary
stars, respectively.
{\it Circles} represent REOSC-echelle spectra (bigger symbols
refer to data used in the calculation of the orbital solution);
{\it squares} stand for
B\&C and REOSC-DS spectra and {\it triangles}, for CITS spectra.
Typical error bars for each data-set are also indicated.}
\includegraphics{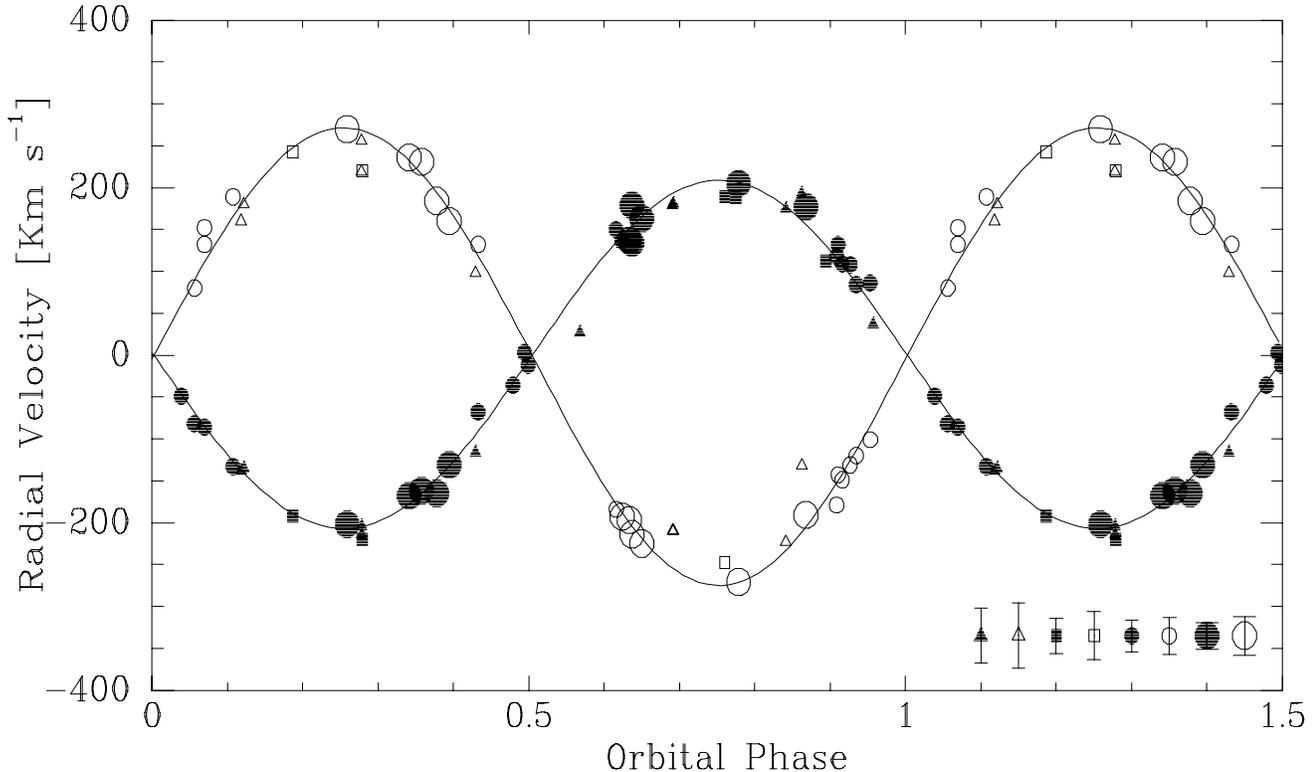}
\label{vr}
\end{figure*}

Figure~\ref{vr} represents the complete set of observed radial 
velocities as a function of the adopted
orbital period, along with the circular orbital solution from Table 2.

\begin{table}
\label{symbols}
\caption{Circular orbital elements of CPD $ -59^{\circ}$\,2635.}
\begin{tabular}{lc} 
\\
$P$ [days] & 2.29995  $\pm$ 2 $\times 10^{-5}$ \\
$K_1$ [$km\,s^{-1}$] & 208 $\pm$ 2 \\
$K_2$ [$km\,s^{-1}$] & 273 $\pm$ 2 \\
$\gamma$ [$km\,s^{-1}$] & 0 $\pm$ 1 \\
$T_{max}$ [HJD] & 2450845.664  $\pm$  0.01 \\
$a_1.sin i$ [$R_{\odot}$] & 9.4 $\pm$ 0.1 \\
$a_2.sin i$ [$R_{\odot}$] & 12.4 $\pm$ 0.1 \\
$M_1\,sin^3 i $ [$M_{\odot}$] & 15.0 $\pm$ 0.5 \\
$M_2\,sin^3 i $ [$M_{\odot}$] & 11.4 $\pm$ 0.5 \\
$q(M_2/M_1)$ & 0.76 $\pm$ 0.01 \\
\end{tabular}

 \medskip
 NOTE: $T_{max}$ represents the time of maximum radial velocity of
 the primary component.
\end{table}

\subsection{Spectral classification of the binary components
of CPD\,-59$^{\circ}$\,2635.}

CPD\,-59$^{\circ}$\,2635 has been previously classified by Walborn (1982) 
as O7Vnn; 
and by Levato \& Malaroda (1982) as O8/9:V ``+ companion ?'',
already pointing out its probable binary nature.
Massey \& Johnson (1993) classified CPD\,-59$^{\circ}$\,2635 as O8.5V.
The spectrum of CPD\,-59$^{\circ}$\,2635 (identified as star number 516),
shown in the last mentioned paper displays indeed double lines,
apparently not noticed by the authors.

\begin{figure*}
 \vspace{95mm}
 \caption{Spectra of CPD -59$^{\circ}\,2635$ obtained 
at CASLEO with the B\&C spectrograph. 
Spectral features marked are:
 H Balmer lines,
 He{\sc i}(+{\sc ii})~4026 \AA, He{\sc i}~4143~\AA, He{\sc ii}~4200~\AA,
 He{\sc i}~4388~\AA, He{\sc i}~4471~\AA, He{\sc ii}~4541~\AA,
 C{\sc iii}~+~O{\sc ii}~4650~\AA\ blend, He{\sc ii}~4686~\AA, 
 He{\sc i}~4713~\AA, and He{\sc i}~4921~\AA. 
 Note the difference in relative intensities of He{\sc i}~4471\AA\ 
 and He{\sc ii}~4541\AA\ between both spectra.}
\includegraphics{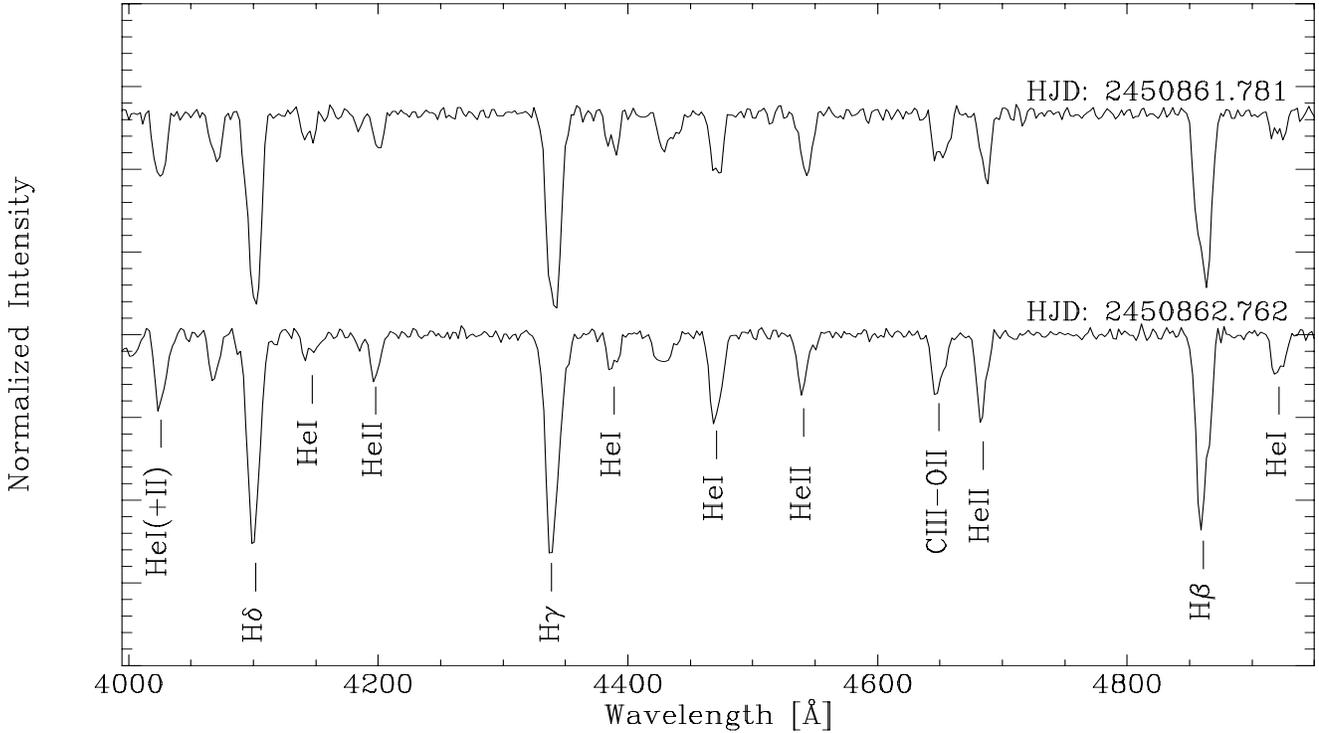}
\label{dslines}
\end{figure*}

Two spectra of \cpd\ from our lower resolution CCD images, corresponding 
to nearly opposite binary phases, are illustrated 
in Figure~\ref{dslines}, where a difference in the relative intensities
 of He{\sc i} 
4471\AA\ and He{\sc ii} 4542\AA\ is evident. In the upper spectrum shown
in Figure~\ref{dslines}, obtained at binary phase 0.76P,
 He{\sc i} and He{\sc ii} 
absorptions appear similar indicating a spectral type O7;
while in the lower spectrum, obtained at the binary phase 0.18P, He{\sc i}
absorption is clearly stronger than that of He{\sc ii}, corresponding to
a spectral type O8-9.
Such spectral variations might explain the slightly different classifications
given to this star by different authors. 

Keeping in mind the known difficulties in classifying spectra of 
close binaries, and the above illustrated spectral variations, 
we have nevertheless tried to estimate the spectral types 
of the binary components of  CPD -59$^{\circ}$\,2635. 

An inspection of our high and intermediate resolution observations 
showed that both stars present absorption-line ratios
of He{\sc i}/He{\sc ii} $\geq$ 1, indicating spectral types 
probably later than O7.
From the spectra with maximum separation of the lines of the binary 
components, and using the classification criteria described by 
Walborn \& Fitzpatrick (1990), we derived spectral types of O8:V and 
O9.5:V for the primary and secondary components, respectively.

In order to provide an additional estimate for the spectral types
of the binary components of CPD -59$^{\circ}$\,2635, we considered the 
He{\sc i}(4922)/He{\sc ii}(5411) ratio, following
the classification criteria proposed by Kerton et al. (1999).
The equivalent widths ($W$) of the 5411 \AA\ and 4922 \AA\
lines were measured in normalized spectra, by means of a Gaussian fit 
to the absorption profiles, though He{\sc ii} 5411 \AA\ looks
somewhat blended even at phases of maximum radial velocity separation
(as above discussed) which causes less confident $W$ determinations. 
We found equivalent width  ratios R($W_{4922}/W_{5411}$) of 0.54 $\pm$ 0.03
and 1.18 $\pm$ 0.04 for the primary and secondary components, respectively,
corresponding to spectral types of O8V and O9.5V, with some variations
depending probably on binary phase.
  
\begin{figure}
\vspace{85mm}
\includegraphics{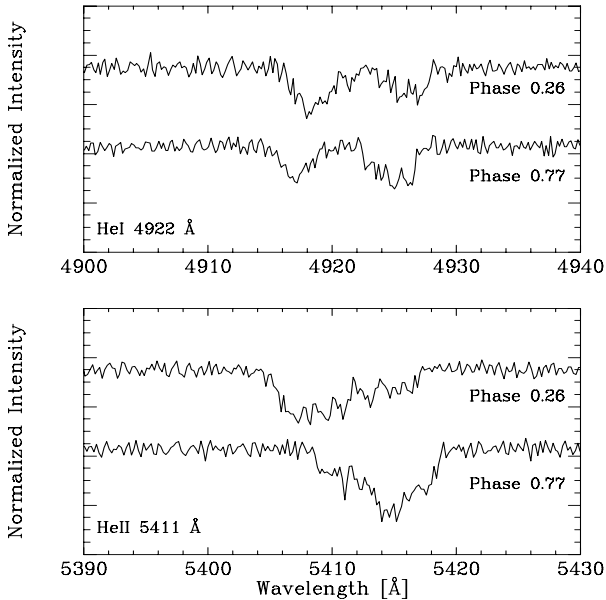}
\caption{He{\sc i} 4922 \AA\ and He{\sc ii} 5411 \AA\ lines
in the spectrum of CPD -59$^{\circ}$\,2635 at phases 0.26 and 0.77}
\label{clasif}
\end{figure}

Figure~\ref{clasif} shows the spectral region comprising 
 He{\sc i} 4922 \AA\ and He{\sc ii} 5411 \AA\ lines
on echelle spectrograms of  CPD\,-59$^{\circ}$\,2635
near the phases of maximum separation of the components.

By inspection of the He{\sc i} and He{\sc ii}  
spectral lines observed during different phases of the binary motion,
we found line depth variations that can be appreciated in
 Figures~\ref{dl4686} and ~\ref{clasif}.
We identify this phenomenon as the
{\it Struve-Sahade effect} (cf. Gies et al. 1997 and references therein) 
observed in massive close binaries, which consists in the
deepening of the spectral lines of the secondary star when it approaches
to the observer.
In our spectra, variable line depths 
seem to be present in both binary components although stronger variations
are observed in the secondary star.




Applying again  the classification criteria proposed by  
Kerton et al. (1999)
we found variations around one subclass in spectral types for
both binary components, going from O7 to O8 for the primary and
O8.5 to O9.5 for the secondary star. However, the errors involved in the $EW$
measurements (especially those affecting the He{\sc ii} 5411 \AA\ line)
are also considerable and, as a consequence, we cannot address any
conclusive statement about the phase dependence of these variations 
until higher resolution and S/N observations are available.


As the binary components of CPD\,-59$^{\circ}$\,2635 are probably
in synchronous rotation with the orbital period, we believe that
the later spectral types (i.e. O8V and O9.5V respectively) 
are  more realistic in describing each star in this binary system,
being the earlier ones probably produced by photo-spheric
heating on each star from its companion and/or the colliding wind region
between the stars.

\subsection{Physical Parameters}

In what follows we will estimate the physical parameters of the 
binary components of \cpd.
We adopt for this star the
visual magnitude and distance modulus of Trumpler 16 published by
Massey \& Johnson (1993), namely $V$ = 9.27 
and $M_v - V_0$ = 12.55 $\pm$ 0.08, close to the
value of $M_{\rm v} - V_0$ = 12.6 obtained by Feinstein et al. (1973).
Also from Massey \& Johnson (1993), we take, for \cpd, 
$E_{B-V}=0.54$, $R=3.2$ and thus $V_0=7.54$.
In order to obtain individual absolute magnitudes, we need an
estimate of the luminosity ratio of the binary components.
We applied the corrected integrated-absorption method of Petrie (1940)
in the way described by Niemela \& Morrison (1988). 
We used equivalent widths  measured for He{\sc i} 4387 and  4471
and He{\sc ii} 4686 \AA\ absorptions in spectra where those
features are better resolved. 
We corrected for continuum overlapping using equivalent widths
of the same lines measured for individual stars
of similar spectral types, taken from Mathys, 1988.
Then we calculated for each selected spectral line the quotient
 (L$_2$/L$_1$=$<(B.K_a)/(A.K_b)>$),
where A is the equivalent width of the line measured in the spectrum 
of the most intense component, 
B is the equivalent width of the same line as observed in the
 weaker component, and
$K_{a,b}$ are the equivalent widths measured for single stars
of spectral types O8\,V and O9\,V,  respectively. 
Performing these measurements in several spectra of \cpd\ observed
near phases of maximum separation of the components we obtained 
an average value of ${\rm L}_2/{\rm L}_1 = 0.45 \pm 0.14$, 
which we have used in the following calculations.
With this luminosity ratio and the adopted distance modulus,
 we obtained individual absolute magnitudes
of $M_{\rm V}$ = -4.61 $\pm$ 0.1 and -3.74 $\pm$ 0.1 
for the primary and secondary components, respectively.

Our classification of the components of CPD\,-59$^{\circ}$\,2635 as O8\,V 
and O9.5\,V corresponds to  $T_{\rm eff}$~=~38450 K 
and  BC~=~-3.68 for the primary,
and $T_{\rm eff}$~=~34620 K and BC~=~-3.36 for the secondary,
according to the calibration of effective temperatures ($T_{\rm eff}$) and
bolometric corrections (BC) proposed by  Vacca, Garmany \& Shull (1996,
hereafter VGS).
However, according to Schmidt-Kaler (1982), the corresponding
effective temperatures
and bolometric corrections would be $T_{\rm eff}$~=~35800 K 
and BC~=~-3.54 for the primary,
and $T_{\rm eff}$~=~31500 K and BC~=~-3.25 for the secondary.  
Therefore, depending on which calibration we adopt, we would find
somewhat different values for the bolometric magnitudes, and thus, 
luminosities of each binary component.

Knowing the luminosities and effective temperatures,
we can derive  the radii of the stars, that we want to compare
with the radii of the critical Roche lobes.
Those were estimated 
using the expression given by Paczynski (1971):

\begin{center}  
$r_{1}\sin{i} = a \sin{i} (0.38 + 0.2~log(M_{1} / M_{2}))$
\end{center}

\noindent
for a ``mean'' Roche radius of $r_{1}$ and separation $a$.
We obtained individual Roche radii of ${\it r}_1 \sin{i}$ = 
8.8 $R_{\odot}$ and ${\it r}_2\sin{i}$ = 7.8 $R_{\odot}$.
We need to know something about the inclination of the orbital plane
in order to compare these critical Roche radii with the
Stefan-Boltzmann radii of the stars.


The physical parameters derived for the binary components are summarized
in Table~\ref{parametros}.

\begin{table*}
\caption{Physical Parameters of CPD -59$^{\circ}$\,2635.}
\label{parametros}
\begin{tabular}{l|cc||cc} 

  &\multicolumn{2}{c||}{Primary component} & \multicolumn{2}{c||}
{Secondary component} \\
Calibration & VGS    & Schmidt-Kaler & VGS & Schmidt-Kaler\\
                   \\ 
$T_{\rm eff}$ [$K$]  & 38450  & 35800 & 34620  & 31500  \\
$M_{\rm bol}$& -8.3 $\pm$ 0.15 & -8.1 $\pm$ 0.15 
& -7.1 $\pm$ 0.15 & -7.0 $\pm$ 0.15 \\
$L$ [$L_{\odot}$] &162000 $\pm$ 20000& 143000 $\pm$ 20000 
& 54000 $\pm$ 11000 & 49000 $\pm$ 11000  \\
$R_{\rm S-B}$  [$R_{\odot}$]& 9.1 $\pm$ 0.4 
& 9.8 $\pm$ 0.4 & 6.5 $\pm$ 0.5 & 7.4 $\pm$ 0.5\\ 
R$_{\rm S-B}$/R$_{\rm Roche-lobe}$&0.8 $\pm$ 0.1 &0.9 $\pm$ 0.1
&0.7 $\pm$ 0.1 &0.8 $\pm$ 0.1 \\
$V_{\rm rot}$ [$km.s^{-1}$]& 200 $\pm$ 10 & 215 $\pm$ 10 & 142 $\pm$ 15
& 164 $\pm$ 15 \\
\end{tabular}

 \medskip
Notes:
~a) R$_{\rm S-B}$ means Stefan-Boltzmann radii.~~~~~~~~~~~~~~~~~~~~~~~~~~~~~~~~~~~~~~~~~~~~~~~~~~~~~~~~~~~~~~~~~~~~~~~~~~~~~~~~~~~~~~~~~~~~~~~~~~~~~~~~~~~~~

~~~~~~~~~b) Theoretical Roche-lobe radii were desafected by the $\sin i$ factor using a probable average value for the inclination of the orbital plane, namely 60$^{\circ}$ .~~~~~~~~~~~~~~~~~~~~~~~~~~~~~~~~~~~~~~~~~~~~~~~~~~~~~~~~~~~~~~~~~~~~~~~~~~~~~~~~~~~~~~~~~~~~~~~~~~~~~~~~~~~~
\end{table*}

Under the assumption that the system is in synchronous rotation
(which seems reasonable in a massive binary of short period)
we derived probable rotational velocities for its components
(quoted in Table~\ref{parametros}
which are higher than the observed rotational
velocities of single stars of similar spectral types
(e.g. Slettebak et al., 1975 ; Conti \& Ebbets, 1977). 

We tried to estimate the projected rotational velocities
($V\, sin\,i$)  comparing the observed absorption profiles
of He{\sc i} 4388~\AA\ and 4471~\AA\  with flux profiles from non-LTE
model atmospheres by Auer \& Mihalas (1972). We chose models 
corresponding to $T_{\rm eff}$ of 40000, 35000 and 30000~K and
$log\,g$ = 4 to represent the binary components.
The model profiles were convolved with different rotational velocity profiles, 
finding satisfactory agreement with observations for $V\,sin\,i$ values of
180 $\pm$ 25 km\,s$^{-1}$ and 140 $\pm$ 30 km\,s$^{-1}$ for the
primary and secondary components, respectively. 
Comparing these results with the calculated rotational
velocities, we obtain, for the inclination
of the orbital plane, values of 64$^{\circ}$ and 79$^{\circ}$, 
for the primary and secondary, respectively, 
using the radii derived through the  calibration by VGS, or
56$^{\circ}$ and 59$^{\circ}$, for primary and secondary, respectively,
using the calibration by Schmidt-Kaler (1982).
However, the errors involved in the $V \sin i$ determination give
room for a large range of inclinations.


Assuming the mass-spectral type relation for normal O-type stars 
by VGS (1996), we can expect masses near 25 and
21 $M_{\odot}$ for the individual binary components of 
CPD\,-59$^{\circ}$\,2635.
These values are similar to (or slightly larger than)
 those obtained from the observation
of eclipsing binary systems with O8V and O9V components
(e.g. EM Car, Andersen \& Clausen, 1989; Y Cyg, Burkholder, Massey
\& Morrell, 1997; CQ Cep, Kartascheva \& Svechnikov, 1989).
 This also points to an orbital inclination  inclination
of 56$^{\circ}$ $\pm$ 6$^{\circ}$, similar to the values estimated
from the line widths. 
If these estimates are correct, the system is not likely to present
eclipses. However, some kind of light variations may occur
due to tidal deformation, considering that both components 
are hot luminous stars in a close binary system.
Also, if our guess for the inclination is as supposed, the system
 would be detached, with both components within their critical Roche
lobes, as we would expect for a young binary with non evolved components.

\section{X-ray Emission}

CPD $-59^{\circ}$ 2635 was detected as a serendipitous X-ray source during
numerous pointing in the Carina Nebula by the \rosat\ X-ray satellite
observatory with both the Position Sensitive Proportional Counter
(PSPC) and the High Resolution Imager (HRI).  In the PSPC images the
star is unresolved from other nearby X-ray sources (notably the O
stars HD 93343 and Tr 16 \#182) due to the rather coarse spatial
resolution of the PSPC ($\sim 1'$).  The HRI has finer spatial
resolution ($\sim 10''$) and so can better resolve CPD 
$-59^{\circ}$ 2635 from
surrounding sources, providing a more accurate measure of the
uncontaminated X-ray emission from the star.
Table 4 lists 3 deep \rosat\ HRI pointing which include 
CPD $-59^{\circ}$ 2635.\\

The source lies about
$ 4.2' $ off axis, and at this position the 50\% encircled energy radius
is about $3''$.  We extracted source counts from these 3 HRI
sequences in an $8''$ region centered on CPD $-59^{\circ}$ 2635.  We extracted
background counts from a region of blank sky centered at
$\alpha_{2000} = 10^{h}$ $45^{m}$ $19.6^{s}$, $\delta_{2000}=
-59^{\circ}$ $44'$ $43.2''$ using an extraction radius of $40''$ to
reduce the statistical uncertainty in the net rate.\\

\begin{table}
\caption{Three deep \rosat\ HRI pointings which include 
CPD $-59^{\circ}$\,2635.}
\begin{tabular}{lccc} 
\\
Sequence Identification & Begin Date & End Date & Exposure\\ 
\\
RH900385N00   & 1992-07-31 & 1992-08-02 & 11527   \\
RH900385A03   & 1994-07-21 & 1994-07-29 & 40555   \\
RH202331N00   & 1997-12-23 & 1998-02-10 & 47095   \\
\end{tabular}
\end{table}

\begin{figure} 
\vspace{65mm}
\caption{HRI X-ray light curve of CPD $-59^{\circ}$\,2635.}
\includegraphics{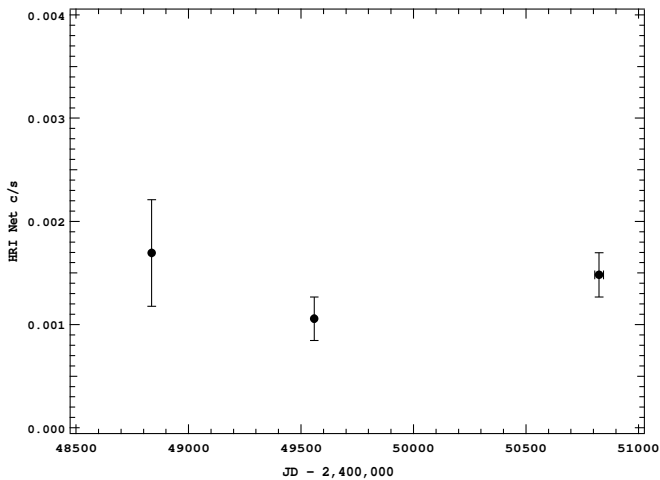}
\label{cpd_lightcurve.eps}
\end{figure}

Figure~\ref{cpd_lightcurve.eps} shows the extracted net light curve for the source.  There is
no evidence for variability.  Because the HRI has no spectral
resolution, we could not determine X-ray luminosity by direct spectral
fitting.  Rather, to determine the X-ray luminosity we converted the
total net count rate to luminosity by assuming a Raymond-Smith thermal
source spectrum with a temperature $kT = 0.5$ keV and an absorbing
column $N_{H}= 2 \times 10^{21}$ cm$^{-2}$, which should reasonably
approximate the X-ray emission from OB type stars in the Carina
nebula.  At $4'$ off-axis, the HRI vignetting correction is 1.01
(David, et al., 1997); the total net count rate, corrected for
vignetting, is $1.35\pm 0.14 \times 10^{-3}$ HRI counts s$^{-1}$.
With our assumed spectral parameters, this corresponds to an X-ray
luminosity $L_{x} = 3.8\times 10^{31}$ ergs s$^{-1}$ in the \rosat\
band ($0.2-2.4$ keV), uncorrected for absorption, using $M_{v} - V_{0}
= 12.55$.  After correcting for absorption, the X-ray luminosity at
the source is approximately $L_{x, unabs} = 9 \times 10^{31}$ ergs
s$^{-1}$ in the \rosat\ band.  
With a total bolometric luminosity $L_{tot} \sim  200000 L_{\odot}$ 
the ratio of the unabsorbed X-ray luminosity  to the total 
luminosity is $\log L_{x}/L_{tot} \sim -7$,  
similar to the  canonical value of this ratio in the $ROSAT$ band
(Bergh\"offer et al.  1997). 
According to this, \cpd\ does not show excess in its X-ray flux
compared to other O-type stars, and the question about why 
it looks brighter than its neighbour HD~93343 in the $ROSAT$ HRI
image of Figure~\ref{hri_dss_cmp.eps} remains to be clarified when
 more observations of both stars are available.

\section{Conclusions}

In the present study we demonstrate that 
CPD\,-59$^{\circ}$\,2635 is a close binary system in a circular orbit.
Both binary components are O-type stars, and the short period of binary
motion suggests strong interactions between
the stellar winds.

Variations of the order of one subclass in the spectral types of
both components are observed, probably related to the phenomenon
known as
 {\it Struve-Sahade effect}. 
A possible explanation for the Struve-Sahade effect,
analyzed by Gies et al. (1997), is that it is present in
systems expected to contain colliding winds with
X-ray generation from the bow shock between the stars. 
However, the observed X-ray light curve does not show any significant
variations, and moreover, the ratio $L_x / L_{bol}$
seems to be similar to that observed for other O-type stars.


This star is a  massive close binary, with hot, luminous components 
(O8V + O9.5V) and thus a good candidate for 
further exploration of colliding wind effects.

With the estimated inclination $\sim$~60$^{\circ}$, we can conclude that
the stellar radii are smaller than the corresponding Roche lobes,
the system being detached, as one would expect from the evolutionary
status of a member in a cluster still containing unevolved O3 stars.

Though the system is not expected to present eclipses,
future photometric studies could reveal the presence of 
phase locked light variations produced by tidal deformation of the
binary components.
This would provide the opportunity of making a better estimate of the
 inclination of 
the orbital plane, and consequently lead to the derivation of 
absolute individual masses for the  components of \cpd.
No need to recall that this is of fundamental astrophysical interest
concerning O-type stars.

\section{Acknowledgements}
We thank the director and staff of CASLEO for
technical support and kind hospitality during the observing runs.
N.I.M. is indebted to the director and staff of CTIO for the use of
their facilities.
We acknowledge use at CASLEO of the CCD and data acquisition system
supported under U.S. NSF grant AST-90-15827 to R. M. Rich.\\
The Digitized Sky Survey in the southern hemisphere is based on
photographic data obtained using The UK Schmidt Telescope.  The UK
Schmidt Telescope was operated by the Royal Observatory Edinburgh,
with funding from the UK Science and Engineering Research Council,
until 1988 June, and thereafter by the Anglo-Australian Observatory.
Original plate material is copyright the Royal Observatory Edinburgh
and the Anglo-Australian Observatory.  The plates were processed into
the present compressed digital form with their permission.  The
Digitized Sky Survey was produced at the Space Telescope Science
institute under US Government grant NAG W-2166.  This research has
made use of NASA's Astrophysics Data System Abstract Service.  This
research has made use of data obtained from the High Energy
Astrophysics Science Archive Research Center (HEASARC), provided by
NASA's Goddard Space Flight Center.\\
Useful discussions with R. Barb\'a are very much appreciated.

\addcontentsline{toc}{section}{Reference}

\end{document}